\documentclass[aps,twocolumn,prd,nofootinbib]{revtex4}
\usepackage{txfonts}
\usepackage{graphicx}
\usepackage{dcolumn}
\usepackage{bm}
\usepackage{amssymb}
\usepackage{latexsym}

\newcommand{\be}{\begin{equation}}
\newcommand{\ee}{\end{equation}}
\newcommand{\bq}{\begin{eqnarray}}
\newcommand{\eq}{\end{eqnarray}}

\begin{document}


\title{Can black holes be torn up by phantom dark energy in cyclic cosmology?}

\author{Xin Zhang}
\affiliation{Department of Physics, College of Sciences,
Northeastern University, Shenyang 110004, People's Republic of
China} \affiliation{Kavli Institute for Theoretical Physics China at
the Chinese Academy of Sciences, P.O.Box 2735, Beijing 100080,
People's Republic of China}

\begin{abstract}
Infinitely cyclic cosmology is often frustrated by the black hole
problem. It has been speculated that this obstacle in cyclic
cosmology can be removed by taking into account a peculiar cyclic
model derived from loop quantum cosmology or the braneworld
scenario, in which phantom dark energy plays a crucial role. In this
peculiar cyclic model, the mechanism of solving the black hole
problem is through tearing up black holes by phantom. However, using
the theory of fluid accretion onto black holes, we show in this
paper that there exists another possibility: that black holes cannot
be torn up by phantom in this cyclic model. We discussed this
possibility and showed that the masses of black holes might first
decrease and then increase, through phantom accretion onto black
holes in the expanding stage of the cyclic universe.

\end{abstract}


\maketitle

The oscillating or cyclic model of the universe is an attractive
idea in theoretical cosmology since it provides the universe with an
infinite life satisfying, in some sense, human being's philosophical
psychology of expecting eternalness. In cyclic cosmology, the
universe oscillates through a series of expansions and contractions.
The idea of an oscillating universe was first proposed by Tolman in
the 1930's \cite{Tolman}. In recent years, Steinhardt, Turok and
collaborators \cite{cycsteinhardt} proposed a cyclic model of the
universe as an alternative to the inflation scenario, in which the
cyclicity of the universe is realized in the light of two separated
branes. In general, however, cyclic universe models confront two
severe problems making the infinite cyclicity impossible. First, the
black holes produced in the universe, which cannot disappear due to
the Hawking area theorems, grow ever larger during subsequent
cycles, and they eventually will occupy the entire horizon volume
during a contracting phase so that calculations in cyclic models
break down. The second problem is that the entropy of the universe
increases from cycle to cycle due to the second law of
thermodynamics, so that extrapolation into the past will lead back
to an initial singularity.

Recently, a new version of oscillating cosmology
\cite{Brown:2004cs,Baum:2006nz} (also dubbed ``phantom bounce'' in
\cite{Brown:2004cs}) claimed that the problems of black holes and
entropy puzzling cyclic models can be resolved by means of the
peculiar characteristic of the phantom dark energy in the universe.
Usually, the phantom energy density becomes infinite in a finite
time, leading to the big-rip singularity \cite{Caldwell:2003vq}.
However, we expect that an epoch of quantum gravity sets in before
the energy density reaches infinity. Therefore, we arrive at the
notion that quantum gravity governs the behavior of the universe
both at the beginning and at the end of the expanding universe,
where the energy density is enormously high. The high energy density
physics may lead to modifications to the Friedmann equation; for
example, it may introduce a negative $\rho^2$ term, such as in loop
quantum cosmology \cite{loop} and braneworld scenario
\cite{Shtanov:2002mb}, which causes the universe to bounce when it
is small, and to turn around when it is large. The cyclic scenario
discussed in this paper is distinguished from the Steinhardt$-$Turok
cyclic scenarios in that the phantom energy plays a crucial role. In
such a cyclic cosmology, it was shown in \cite{Brown:2004cs} that
black holes in the universe will be torn up by the phantom dark
energy before the turnaround. Also, in \cite{Baum:2006nz} the
authors claimed that at the turnaround of the cyclic cosmology both
volume and entropy of our universe will decrease by a gigantic
factor, while very many independent similarly small contracting
universes will be spawned, thus resolving the entropy problem of the
cyclic cosmology.\footnote{For a different viewpoint on this
scenario, see \cite{Zhang:2007du}.}

The key idea for eliminating the black hole problem in cyclic
cosmology is that any black holes formed in an expanding phase of
the universe are torn apart by the phantom dark energy before they
can create problems during contraction. However, the discussions on
destruction of black holes in \cite{Brown:2004cs} are based on a
rough evaluation. In general relativity, the source for a
gravitational potential is the volume integral of $\rho+3p$.
Therefore, an object of radius $R$ and mass $M$ is pulled apart when
$-(4\pi/3)(\rho+3p)R^3\sim M$. A black hole with radius $R=2GM$ is
thus torn up when the phantom energy density has climbed up to a
value $\rho_{\rm bh}\sim (3/32\pi)(M^2G^3|1+3w|)^{-1}$, where $G$ is
the Newton gravitational constant, and $w=p/\rho<-1$ is the equation
of state of the phantom dark energy. For ensuring that the black
holes are destroyed before turnaround, one only needs $\rho_{\rm
bh}<\rho_{\rm c}$, where $\rho_{\rm c}$ is the critical energy
density in the cyclic model, namely the energy density corresponding
to the turnaround (and bounce).

However, the destruction of black holes by phantom accretion is not
an instantaneous behavior, it is a process actually. In a phantom
dominated universe with ``big rip'', black holes can be torn up
completely by phantom energy before the big rip, however, in such a
cyclic cosmology caused by the ``phantom bounce'', whether or not
black holes can be torn up by phantom energy should be investigated
in detail by using the theory of dark energy accretion by black
holes. We shall study in this paper the phantom accretion onto a
black hole in the cyclic universe (in the expanding phase dominated
by phantom component), and show that there exists the possibility
that the mass of the black hole may decrease first, to a minimum,
and increase then, until restoring the original mass value at the
turnaround. Thus, actually, according to this possibility, phantom
energy cannot help resolve the black hole problem in cyclic
cosmology.

For the fluid accretion onto a black hole, Babichev et al. have
obtained a successful mechanism \cite{Babichev:2004yx} in which, as
a consequence of fluid accretion, the mass of the black hole changes
at a rate $\dot{M}=4\pi AM^2[\rho_\infty+p(\rho_\infty)]$, where $A$
is a positive dimensionless constant, and $\rho_\infty$ and
$p(\rho_\infty)$ are the energy density and pressure of the fluid at
the remote distance from the black hole, respectively. For the case
of a wormhole, see \cite{GonzalezDiaz:2004vv}. Following the
mechanism of Babichev et al., we shall study the phantom accretion
of black hole in the cyclic cosmology in this paper.

Let us consider a modified Friedmann equation in which a $\rho^2$
term with negative sign is introduced due to some quantum gravity
effects,
\begin{equation}
H^2={8\pi G\over 3}\rho\left(1-{\rho\over\rho_{\rm
c}}\right),\label{modiFeq}
\end{equation}
where $H=\dot{a}/a$ is the Hubble parameter, and $\rho_{\rm c}$ is
the critical energy density set by quantum gravity, distinguished
from the usual critical density $3M_{\rm pl}^2H^2$ (where $M_{\rm
pl}=1/\sqrt{8\pi G}$ is the reduced Planck mass). This modified
Friedmann equation can be derived from the effective theory of loop
quantum cosmology \cite{loop}, and also from the braneworld scenario
\cite{Shtanov:2002mb}. In loop quantum cosmology, the critical
energy density can be evaluated as $\rho_{\rm c}\approx
0.82\rho_{\rm pl}$, where $\rho_{\rm pl}=G^{-2}=2.22\times
10^{76}~{\rm GeV}^4$ is the Planck density. In the braneworld
scenario, $\rho_{\rm c}=2\sigma$, where $\sigma$ is the brane
tension, and a negative sign in Eq. (\ref{modiFeq}) can arise from a
second timelike dimension but that gives difficulties with closed
timelike paths. In models motivated by the Randall$-$Sundrum
scenario \cite{Randall:1999}, the most natural energy scale of the
brane tension is of the order of the Planck mass, but the problem
can be generally treated for any value of $\sigma>{\rm TeV}^4$.

While we use loop quantum cosmology only as an example, we feel that
it would be better to make some clarifications for the background
for avoiding misunderstandings. For instance, the value for $\rho_c$
given above is based on an ad hoc choice and not derived from a
general setting. In general, there can be a different numerical
factor, and the critical density can even depend on the scale factor
depending on the precise underlying state \cite{Bojowald:2008ec}.
Moreover, the evolution equation in the form (\ref{modiFeq})
provides the correct effective theory only in the case of a stiff
fluid (such as a free, massless scalar) which, strictly speaking, is
not the case discussed here. In general, there are additional
quantum corrections due to the interacting nature of the system. The
latest status on this issue can be seen, e.g., in
\cite{Bojowald:2008ec}. Therefore, it should be stressed that the
simple equation (\ref{modiFeq}) only serves as an example, and one
should not indiscreetly say that this is the general situation.

In the regime very near the turnaround, the term $(1-\rho/\rho_c)$
in (\ref{modiFeq}) is almost zero, and one can expect that
additional correction terms become very important. This is true for
loop quantum cosmology, and for braneworlds one might similarly
expect additional corrections due, e.g., to higher order terms in
the string tension. Therefore, although the equation (\ref{modiFeq})
is frequently used in the literature, the readers should be
cautioned that the equations themselves might not give the full
picture. In this paper, however, we only consider the simple case
based on Eq. (\ref{modiFeq}), i.e., we treat the problem
phenomenologically, following the literature, such as
\cite{Brown:2004cs,Baum:2006nz}.

Such a modified Friedmann equation with a phantom energy component
leads to a cyclic universe scenario in which the universe oscillates
through a series of expansions and contractions
\cite{Brown:2004cs,Baum:2006nz}. Phantom energy can dominate the
universe today and drive the current cosmic acceleration
\cite{phantom}. Then, as the universe expands, it becomes more and
more dominant and its energy density becomes very high. When the
phantom energy density reaches the critical value $\rho_{\rm c}$,
the universe reaches a state of maximum expansion which we call
``turnaround'', and then begins to recollapse, according to the
modified Friedmann equation. The contraction of the universe makes
the phantom energy density dilute away and the matter density
dominate. Once the universe reaches its smallest extent, the matter
density hits the value of the critical density, the modified
Friedmann equation leads to a ``bounce'', making the universe once
again begin to expand. Note that both turnaround and bounce are
nonsingular in this scenario.

In this paper, we only consider the high energy regime in the
expanding branch, where phantom energy is overwhelming and the
$\rho^2$ effect is prominent. Since in the high energy regime we
have $\rho\gg \rho_{\rm today}$, we say $\rho_{\rm today}\sim 0$.
Phantom dark energy is characterized by the parameter of equation of
state $w=p/\rho<-1$ which is considered to be a constant in this
paper for convenience. Combining the modified Friedmann equation
(\ref{modiFeq}) and the conservation law $\dot{\rho}+3H(\rho+p)=0$
yields
\begin{equation}
\dot{H}=-4\pi(\rho+p)\left(1-{2\rho\over \rho_{\rm
c}}\right),\label{dotH}
\end{equation}
where we have set $G=1$ for convenience (this convention will be
used hereafter). From Eqs. (\ref{modiFeq}) and (\ref{dotH}), we
derive
\begin{equation}
{\ddot{a}\over a}=-{4\pi\over
3}\left\{\rho\left(1-{\rho\over\rho_{\rm
c}}\right)+3\left[p\left(1-{2\rho\over \rho_{\rm
c}}\right)-{\rho^2\over\rho_{\rm c}}\right]\right\}.\label{ddota}
\end{equation}
Comparing to the classical form of the equation, it is convenient to
define the effective energy density and pressure
\begin{equation}
\rho_{\rm eff}=\rho\left(1-{\rho\over\rho_{\rm
c}}\right),\label{rhoeff}
\end{equation}
\begin{equation}
p_{\rm eff}=p\left(1-{2\rho\over\rho_{\rm
c}}\right)-{\rho^2\over\rho_{\rm c}},\label{peff}
\end{equation}
then Eq. (\ref{ddota}) can be written as
\begin{equation}
{\ddot{a}\over a}=-{4\pi\over 3}(\rho_{\rm eff}+3p_{\rm eff}).
\end{equation}
Also, we have $H^2={8\pi\rho_{\rm eff}/ 3}$ and
$\dot{H}=-{4\pi}(\rho_{\rm eff}+p_{\rm eff})$, obviously.

This means that all the quantum gravity effects can be attributed to
the effective quantities such as $\rho_{\rm eff}$ and $p_{\rm eff}$.
According to this perspective, classical general relativity
(Einstein equation) can also be used to describe quantum cosmology,
provided that all the quantum gravity effects could be effectively
attributed to the energy-momentum tensor, i.e. the Einstein equation
under such circumstances should be effectively written as
$G_{\mu\nu}=8\pi T_{\mu\nu}^{\rm eff}$. So, under this description,
the universe looks like filled with the effective fluid with
$\rho_{\rm eff}$ and $p_{\rm eff}$. Using $\rho_{\rm eff}$ and
$p_{\rm eff}$, we can effectively describe the behavior of the
universe.

Given the effective energy density and pressure, the effective
equation of state is defined naturally as
\begin{equation}
w_{\rm eff}={p_{\rm eff}\over \rho_{\rm eff}}={w(1-2x)-x\over
1-x},\label{weff}
\end{equation}
where $x$ is defined as dimensionless density, $x=\rho/\rho_{\rm
c}$, so we have $0<x<1$.

\begin{figure}[htbp]
\begin{center}
\includegraphics[scale=0.9]{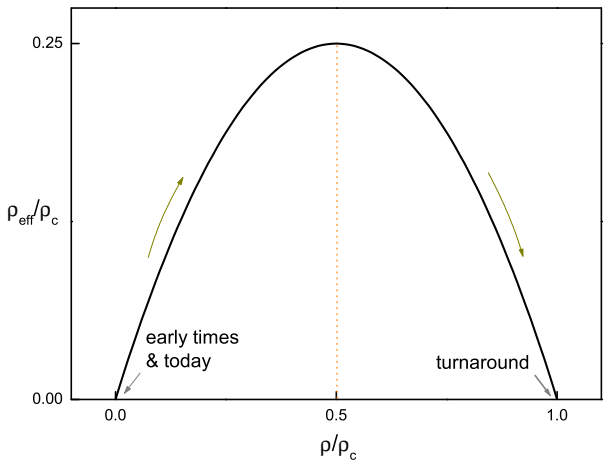}
\caption[]{\small Sketch map of the expanding phase of phantom
dominated universe in the cyclic cosmology. In the expanding stage,
the phantom energy density increases monotonously, whereas the
effective energy density of the universe first increases and then
decreases, implying an effective behavior of ``quintom'', due to the
quantum gravity effects. The effective energy density $\rho_{\rm
eff}$ arrives at its maximum, $\rho_{\rm eff}^{\rm max}=\rho_{\rm
c}/4$, when the phantom density reaches the half value of the
critical density, $\rho=\rho_{\rm c}/2$.}\label{fig:phantomcyc}
\end{center}
\end{figure}

Albeit phantom energy density always increases monotonously with the
expansion of the universe, the effective energy density, however,
exhibits totally different behavior comparing to the phantom energy
density. Fig. \ref{fig:phantomcyc} plots the rewritten Eq.
(\ref{rhoeff}), $y=x(1-x)$, where $y=\rho_{\rm eff}/\rho_{\rm c}$
and $x=\rho/\rho_{\rm c}$. It is clear that the effective energy
density $\rho_{\rm eff}$ first increases and then decreases, which
implies that the effective behavior of the universe under the
quantum gravity domination resembles a ``quintom''\footnote{For
quintom dark energy see \cite{quintom}, and for the detailed
analysis for the effective quintom behavior in loop quantum
cosmology see \cite{Zhang:2007bi}.} whose key feature is that its
equation of state can evolve across the ``cosmological constant
boundary''. One can check that $w_{\rm eff}<-1$ in the range
$0<x<1/2$, and $w_{\rm eff}>-1$ within $1/2<x<1$, provided that
$w<-1$. Hence, we learn that the place $x=1/2$ plays the role of the
``phantom divide'' for the effective energy density of the universe.
The dimensionless effective energy density $y$ arrives at its
maximum, $y_{\rm max}=1/4$, when the dimensionless phantom density
reaches the value $x=1/2$.

\begin{figure}[htbp]
\begin{center}
\includegraphics[scale=0.9]{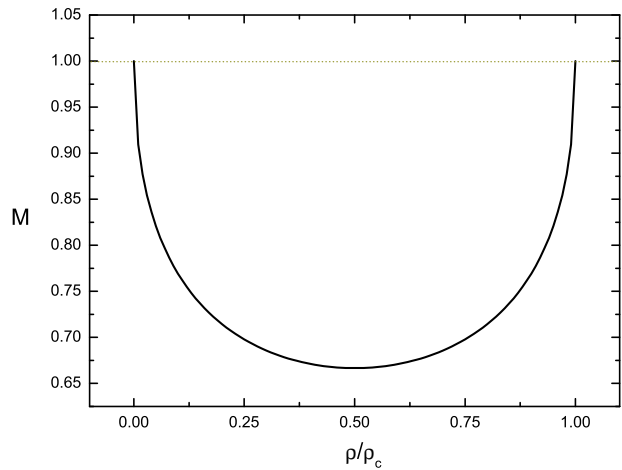}
\caption[]{\small The variation of black hole mass due to the
phantom dark energy accretion in the cyclic cosmology. Here we show
a simplified case, $M(x)=1/(1+\sqrt{x(1-x)})$, where
$x=\rho/\rho_{\rm c}$, for illustration. In the cyclic universe
dominated by phantom dark energy, the phantom accretion makes the
black hole mass first decrease, to a minimum, then increase, until
restoring its initial value at the turnaround.}\label{fig:bhmass}
\end{center}
\end{figure}

Consider now the phantom dark energy accretion of a Schwarzschild
black hole in such a cyclic universe. As shown above, at the level
of effective theory, the quantum gravity effects make the phantom
energy behave as a quintom energy in such a universe. Thus, at
remote distances from the black hole (namely, on the cosmic scale),
the universe looks like filled with the effective fluid with
$\rho_{\rm eff}$ and $p_{\rm eff}$. That is to say, $\rho_\infty$
behaves as $\rho_{\rm eff}$ when taking the quantum gravity effects
into account. Hence, considering the resulting cosmological
evolution in this model, we should replace $\rho_\infty$ by
$\rho_{\rm eff}$ (and replace $p(\rho_\infty)$ by $p_{\rm eff}$).
However, in the vicinity of the black hole (namely, on the local
scale), we assume that the dark energy is still described by $\rho$
and $p$ (note that here $\rho$ and $p$ are not quantities on the
cosmic scale, so they are not homogeneous due to the influence of
the black hole), since the physics on cosmic scale does not directly
affect the physics on local scale. Following Babichev et al.
\cite{Babichev:2004yx}, we can deal with the fluid accretion onto a
black hole. First, the integration of the time component of the
energy-momentum conservation law $T^{\mu\nu}_{;\nu}=0$ gives the
first integral of motion for the stationary spherically symmetric
accretion:
\begin{equation}
(\rho+p)\left(1-{2M\over r}+u^2\right)^{1/2}{u r^2\over
M^2}=B,\label{1stint}
\end{equation}
where $u=dr/ds$, $r$ is the Schwarzschild radial coordinate, $M$ is
the mass of a black hole, and $B$ is a constant. Next, using the
projection of the energy-momentum conservation law on the
four-velocity $u_\mu T^{\mu\nu}_{;\nu}=0$, where $u^\mu=dx^\mu/ds$
is the fluid four velocity with $u^\mu u_\mu=1$, one obtains the
second integral of motion:
\begin{equation}
{u r^2\over M^2}\exp\left[\int_{\rho_\infty}^\rho {d\rho'\over
\rho'+p(\rho')}\right]=-A,\label{2ndint}
\end{equation}
where $u<0$ stands for the case of inflow motion and $A$ is a
positive dimensionless constant. From (\ref{1stint}) and
(\ref{2ndint}), one can easily derives
\begin{equation}
\begin{array}{l}
(\rho+p)\left(1-{2M\over
r}+u^2\right)^{1/2}\exp\left[\int_\rho^{\rho_\infty} {d\rho'\over
\rho'+p(\rho')}\right]\\=\rho_\infty+p(\rho_\infty).\end{array}\label{12int}
\end{equation}
The fluid accretion gives rise to that the mass of a black hole
changes at a rate $\dot{M}=-4\pi r^2 T^r_0$. With the help of
(\ref{2ndint}) and (\ref{12int}), one obtains \cite{Babichev:2004yx}
\begin{equation}
\dot{M}=4\pi AM^2[\rho_\infty+p(\rho_\infty)].\label{mass}
\end{equation}
This shows that the rate of change of the black hole mass is
determined by the behavior of dark energy on cosmic scales.

As mentioned, the cosmological evolution behavior of phantom energy
in the cyclic cosmology mimics the behavior of a quintom energy due
to the modified Friedmann equation $H^2\sim \rho_{\rm eff}$. Thus,
it is obvious that $\rho_\infty$ and $p(\rho_\infty)$ in Eq.
(\ref{mass}) should be replaced by $\rho_{\rm eff}$ and $p_{\rm
eff}$.\footnote{This point can be easily understood. In
\cite{Babichev:2004yx}, the full formulation of the fluid accretion
onto a black hole is fulfilled within the framework of conventional
general relativity. As aforementioned, the usual general relativity
is also suitable to describe the quantum cosmology provided that all
the quantum gravity effects could be effectively attributed to the
energy-momentum tensor, i.e. $G_{\mu\nu}=8\pi T_{\mu\nu}^{\rm eff}$.
Hence, under such circumstances, the theory of fluid accretion onto
black hole can also be used in the cyclic model of universe provided
that $\rho_\infty$ and $p(\rho_\infty)$ in Eq. (\ref{mass}) are
replaced by $\rho_{\rm eff}$ and $p_{\rm eff}$. By far, we have
shown that the relation $\dot{M}\sim -M^2\dot{H}$ may still hold for
the case of cyclic model.} Therefore, following the procedure of
Babichev et al. \cite{Babichev:2004yx}, one can write down the
change rate of the black hole mass,
\begin{equation}
\dot{M}=4\pi AM^2(\rho_{\rm eff}+p_{\rm eff}).\label{change}
\end{equation}
From this equation, it is clear that the accretion of the phantom
energy with $w<-1$ in a cyclic universe cannot ensure the
monotonously diminishing of the black hole mass since the effective
energy density behaves as a quintom, i.e., $1+w_{\rm eff}$ evolves
from values smaller than $0$ to larger than $0$. Replacing $d/dt$ in
(\ref{change}) with $-\sqrt{3}(1+w)\sqrt{\rho_{\rm eff}}\rho
d/d\rho$, one obtains the following expression,
\begin{equation}
{dM\over M^2}=-{4\pi A(1+w_{\rm eff})\sqrt{\rho_{\rm eff}}\over
\sqrt{3}(1+w)\rho}d\rho.
\end{equation}
By integrating this equation, we obtain
\begin{equation}
M={M_{0}\over 1+CM_{0}},\label{bhvary}
\end{equation}
where $M_{0}$ is the initial mass of the black hole, and $C=8\pi
A\sqrt{\rho_{\rm c}x(1-x)/3}$. Note that here the initial condition
is chosen as $M|_{x_{\rm i}=0}=M_{0}$. Equation (\ref{bhvary})
explicitly indicates that, in the cyclic universe caused by phantom
bounce, through the phantom accretion, black hole mass will decrease
first, and then increase until restoring its initial mass at the
turnaround. The minimum value of the black hole mass, $M_{\rm
min}=M_{0}/(1+4\pi A\sqrt{\rho_{\rm c}/3} M_{0})$, happens at
$x=1/2$. For illustrating the black hole mass variation, we take a
simple case as example, i.e., $M_{0}=1$ and $8\pi A\sqrt{\rho_{\rm
c}/3}=1$, as shown in Fig. \ref{fig:bhmass}. Therefore, so far, we
learn, from the analysis of the fluid accretion of black holes, that
masses of black holes are not diminishing monotonously by phantom
accretion in the cyclic cosmology, i.e., in such a cyclic universe
black holes cannot be torn up by phantom dark energy.

Undoubtedly, in the usual phantom dominated universe with big rip,
phantom accretion onto black holes will always diminish black hole
masses. In this case, the solution of mass variation of black hole
is also in the same form as (\ref{bhvary}) but where $C=8\pi
A\sqrt{\rho/3}$.\footnote{In this case, the phantom dominated
universe has the scale factor as $a(t)=T^{2/[3(1+w)]}$, provided
that $w$ is a constant smaller than $-1$. Here $T=a_{\rm
i}^{3(1+w)/2}+3(1+w)/2 \sqrt{1-\Omega_{\rm m}^0}H_0(t-t_{\rm i})$,
in which $a_{\rm i}$ and $t_{\rm i}$ are the initial values for the
scale factor and time, respectively, at the onset of phantom
domination, and $\Omega_{\rm m}^0$ and $H_0$ are respectively the
fractional matter density and Hubble parameter of today. Hence, for
this case, we have $C(t)=12\pi A(1-\Omega_{\rm
m}^0)H_0^2(|w|-1)(t-t_{\rm i})a_{\rm i}^{3(|w|-1)/2}T^{-1}$.}
Obviously, when the universe goes towards the big-rip,
$\rho\rightarrow \infty$, we have $M\rightarrow 0$, implying that
black holes are torn up by phantom near the big rip. However, in the
cyclic universe discussed in this paper, the phantom dark energy
behaves like a quintom dark energy effectively, due to the quantum
gravity effects, so that the phantom accretion by black holes makes
the black hole masses first decrease and then increase.
Nevertheless, it should also be emphasized that the result obtained
in this paper is heavily based on the work of Babichev et al.
\cite{Babichev:2004yx}.

One may notice, however, that there are some problems in the
calculations of Babichev et al. \cite{Babichev:2004yx}. First, in
their work, the metric used is asymptotically flat (actually, the
Schwarzschild metric) and could not exactly describe the spacetime
of a black hole embedded in a Friedmann$-$Robertson$-$Walker (FRW)
universe. Moreover, Eq. (\ref{mass}) is obtained by ignoring the
backreaction of the phantom matter on the black hole metric. In a
low matter density background, this effect can be safely ignored;
but when the background density becomes large (e.g., comparable to
the black hole density), the metric describing this black hole will
be modified significantly. So, in order to study this issue, in
principle, one should use new exact solutions of the Einstein
equations describing dynamical black holes embedded in an FRW
universe driven by phantom energy and accreting this cosmic fluid.
This issue has been addressed in Ref. \cite{Gao:2008jv}.

In Ref. \cite{Gao:2008jv}, the authors considered the generalized
McVittie metric,
\begin{eqnarray}
ds^2&=&-\frac{\left[1-\frac{M(t)}{2a\left(t\right)r}\right]^2}{
\left[1+\frac{M(t)}{2a\left(t\right)r}\right]^2} \, dt^2
  +{a^2\left(t\right)}\left[1+\frac{M(t)}{2a\left(t\right)r}
  \right]^4
   \nonumber\\ &&\nonumber \\
  &&\times\left(dr^2+r^2d\Omega^2 \right),  \label{metric}
 \end{eqnarray}
in the background of an imperfect fluid with a radial heat flux and,
possibly, a radial mass flow simulating accretion onto a black hole
embedded in an FRW universe. The imperfect fluid is described by the
stress-energy tensor
\begin{equation}
T_{\mu\nu}=\left( p+\rho \right) u_{\mu} u_{\nu} +pg_{\mu\nu} +
q_{\mu} u_{\nu} + q_{\nu} u_{\mu},
\end{equation}
where $u^{\mu}=\left( |g_{00}|^{-1/2}, 0,0,0, \right)$ is the fluid
four-velocity and $q^{\mu}=\left( 0,q,0,0,\right)$ is the radial
heat current. Under this situation, it was shown in
\cite{Gao:2008jv} that the appropriate notion of mass for a
cosmological black hole is the Hawking$-$Hayward quasilocal mass,
\begin{equation}
m_H(t)=M(t)=M_0a(t).
\end{equation}
Obviously, this mass is always increasing in an expanding universe.
In Ref. \cite{Faraoni:2008tx}, the authors further showed that,
under certain assumptions, only those with comoving
Hawking$-$Hayward quasilocal mass are generic, in the sense that
they are late-time attractors. Therefore, we see that if the
backreaction is considered in the calculation of the accretion of
fluid, the conclusion will be opposite to that of Babichev et al.,
i.e., the physical black hole mass may instead increase due to the
accretion of phantom energy. If this is the case, the black hole
will certainly not be torn up by phantom energy even in a usual
phantom dominated universe. This implies that this subject is highly
speculative, and there exist various possibilities that cannot be
excluded or ignored.

In the present work, we are restricted to the framework of Ref.
\cite{Babichev:2004yx}. To avoid misleading readers, we must admit
the drawbacks of this framework, as discussed above. In this
framework, the cosmic fluid, namely the phantom energy, is treated
as a test fluid, so the analysis does not have much to say about the
destruction of black holes in a phantom-dominated universe, which is
a markedly different physical situation. However, actually, we are
currently lacking a satisfactory method to deal with such a subject.
The method in Ref. \cite{Gao:2008jv} is also highly speculative, for
example, the metric solution (\ref{metric}) is also hypothesized but
not derived exactly. Under the circumstances, we can but using the
means currently available to tentatively draw some conclusions on
this topic.

In addition, it should be mentioned that in Ref. \cite{Gao:2008jv}
another interesting possibility was also discussed. In that case,
because the future universe is dominated by phantom dark energy, the
black hole apparent horizon and the cosmic apparent horizon will
eventually coincide and, after that, the black hole singularity will
become naked in finite time, violating the cosmic censorship
conjecture. Therefore, in our case, a question naturally arises
asking whether the cosmological horizon can be smaller than the
black hole horizon during the period of interest. Now we shall
discuss this issue.

In any case we can rewrite Eq. (\ref{bhvary}) as
\begin{equation}
r_{\rm sch}={r^{(i)}_{\rm sch}\over 1+\sqrt{2\pi}AH(t)r^{(i)}_{\rm
sch}},
\end{equation}
where $r^{(i)}_{\rm sch}$ is the initial Schwarzschild radius. The
constant $A$ is of order unity, and in Ref. \cite{Babichev:2004yx}
it is determined, $A=4$. Today the universe is in the low energy
regime, we have $H(t)\sim 0$, so it is clear that $r_{\rm
sch}=r_{\rm sch}^{(i)}$. However, when the universe goes into the
high energy regime, the Hubble parameter $H(t)$ becomes very large;
consequently we have $r_{\rm sch}\sim (\sqrt{2\pi} A)^{-1} H^{-1}$.
In other words, when the Hubble parameter becomes enormously large,
the Hubble radius will be one order of magnitude larger than the
Schwarzschild black hole radius,
\begin{equation}
H^{-1}\sim 10 r_{\rm sch}.
\end{equation}
That is to say, the Hubble volume at least involves about $10^3$
black hole volume at any stages. Note that when $H^{-1}$ diminishes,
$r_{\rm sch}$ also diminishes. In the ordinary phantom cosmology,
there is a ``big-rip'' singularity where $H\rightarrow \infty$ (so
$H^{-1}\rightarrow 0$), so that $r_{\rm sch}\rightarrow 0$ when the
universe approaches the big-rip. This is the essential of Ref.
\cite{Babichev:2004yx}. However, in our case, $H^{-1}$ first
decreases ($0<x<1/2$) and then increases ($1/2<x<1$); so $r_{\rm
sch}$ will also first decrease and then increase until restoring its
initial value $r_{\rm sch}^{(i)}$ at the turnaround point ($x=1$).

Through the above analysis, we show that there may exist another
possibility, i.e., black holes cannot be torn apart by phantom dark
energy in this cyclic universe. The analysis is based on the
assumption that the change rate of black hole mass is determined by
the effective energy density $\rho_{\rm eff}$ and the effective
pressure $p_{\rm eff}$. The reason of considering such a possibility
is due to the inspiration of the models of quintom cyclic universe
(or ``quintom bounce'') \cite{Xiong:2008ic}. Imagining the universe
filled with quintom dark energy with the energy density $\rho$
behaving like the effective quantity $\rho_{\rm eff}$ of this paper,
this quintom matter obviously can cause the cyclicity of the
universe through the usual Friedmann equation $H^2=(8\pi/3)\rho$. In
this situation, through the accretion of quintom matter, the black
hole mass change rate is determined by $\dot{M}=4\pi AM^2(\rho+p)$
indicating that the black hole mass will first decrease and then
increase. The model discussed in this paper is indistinguishable
from the quintom cyclic model \cite{Xiong:2008ic} on the
cosmological evolution. So, inspired by this fact, we consider the
possibility that in the process of fluid accretion onto a black
hole, the effective quantities might play a crucial role.

Of course, we cannot indiscreetly claim that this is a definitive
conclusion. But we cannot exclude this possibility either. Perhaps
one might argue that $\rho_{\rm eff}$ and $p_{\rm eff}$ are not
proper physical quantities, however, the possible mechanism
discussed in this paper provides a novel angle of view to ponder
upon this subject. Taking this view into account, a similar
application in braneworld cosmology has also been discussed in
\cite{MartinMoruno:2007se}.

In summary, we discussed in this paper the problem of whether black
holes can be torn apart by phantom dark energy in cyclic cosmology.
For the cyclic cosmology resting on phantom dark energy described by
(\ref{modiFeq}), it has been viewed that the black hole problem can
be resolved in the light of the characteristic of phantom. However,
we have demonstrated in this paper that there may exist another
possibility. Using the theory of fluid accretion onto black holes,
we analyzed this possibility in detail and showed that the masses of
black holes might first decrease and then increase, through phantom
accretion onto black holes in the expanding stage of the cyclic
universe. Though the conclusion sounds somewhat counterintuitive,
this possibility cannot be excluded. The aim of this paper is to
provide a novel angle of view to this profound topic.

\begin{acknowledgements}

The author acknowledges Rong-Gen Cai, Miao Li, Yun-Song Piao and
Zhi-Guang Xiao for helpful discussions. This work was supported by
the Natural Science Foundation of China (Grant No. 10705041).

\end{acknowledgements}

\end{document}